\documentstyle[twocolumn,aps,prl,graphics,epsfig]{revtex}
\begin{document}

\title{Non-maximally entangled states: production, characterization 
and utilization}
\author{Andrew G. White$^1$, Daniel F. V. James$^2$,
Philippe H. Eberhard$^3$ and Paul G. Kwiat$^1$}
\address{$^1$Physics Division, P-23; $^2$Theoretical Division, T-4;
Los Alamos National Laboratory, Los Alamos, New Mexico 87545}
\address{$^3$Lawrence Berkeley Laboratory, University of California,
Berkeley, California 94720}

\date{Phys. Rev. Lett., to appear; submitted 11 June 1999.}

\draft
\maketitle
\vspace {-3 cm}

\begin{abstract}
Using a spontaneous-downconversion photon source, we produce true 
non-maximally entangled states, i.e., without the need for 
post-selection.  The degree and phase of entanglement are readily 
tunable, and are characterized both by a standard analysis using 
coincidence minima, and by quantum state tomography of the two-photon 
state.  Using the latter, we experimentally reconstruct the reduced 
density matrix for the polarization.  Finally, we use these states to 
measure the Hardy fraction, obtaining a result that is $122 \sigma$ 
from any local-realistic result.
\end{abstract}
\vspace {-0.60 cm}

Entanglement is arguably the defining characteristic of quantum 
mechanics, and can occur between any quantum systems, be they separate 
particles \cite{Schro} or separate degrees of freedom of a single 
particle \cite{Spreeuw}.  The latter can be used to realize 
interference-based all-optical implementations of quantum algorithms 
\cite{optGrover}, while multi-particle entangled states are central in 
discussions of locality \cite{EPR,Bell}, and in quantum 
information, where they enable quantum computation \cite{comp}, 
cryptography \cite{crypto}, dense coding \cite{dense}, and 
teleportation \cite{tele}.  More generally, entanglement is the 
underlying mechanism for measurements on, and decoherence of, quantum 
systems, and thus is central to understanding the quantum/classical 
interface.

Historically, controlled production of multi-particle entangled states 
has proven to be non-trivial.  To date, the ``cleanest'' and most 
accessible source of such entanglement arises from the process of 
spontaneous optical parametric downconversion in a nonlinear crystal 
(for a review, see \cite{2PDC}).  This entanglement is of a specific 
and limited kind: the states are maximally entangled, e.g., 
$(|\rm{HV}\rangle \pm \varepsilon 
|\rm{VH}\rangle)/\sqrt{1+|\varepsilon|^{2}}$, where H and V 
respectively represent the horizontal and vertical polarizations of 
two separated photons, and $\varepsilon=1$.  There is no possibility 
of varying the intrinsic {\it degree of entanglement\/}, $\varepsilon$ 
\cite{entropy}, to produce non-maximally entangled states without 
compromising the purity of the state, i.e., introducing mixture 
\cite{Sak}.  Non-maximally entangled states have been shown to reduce 
the required detector efficiencies for loophole-free tests of Bell 
inequalities \cite{Eberhard}, as well as allowing logical arguments 
that demonstrate the nonlocality of quantum mechanics {\it without\/} 
inequalities \cite{Hardy,cakes,ineqnote}.  More generally, such states 
lie in a previously inaccessible range of Hilbert space, and may 
therefore be an important resource in quantum information 
applications.

States with a {\it fixed\/} degree of entanglement, $\varepsilon 
\simeq 4/3$, have been deterministically generated in ion traps 
\cite{deterministic}, and there have been several optical experiments 
where non-maximally entangled states were controllably generated via 
{\it post-selection\/}, i.e., selective measurement of a product 
state, {\it after\/} the state had been produced 
\cite{HardyExp1,HardyExp2}.  The latter experiments are of 
considerable pedagogical interest in that they demonstrate the logic 
behind inequality-free locality tests.  However, the underlying state 
is factorizable, and so is not truly entangled.  In this letter, we 
describe the controllable production, characterization, and 
utilization of true non-maximally entangled states, generated without 
postselection.

A detailed description of the downconversion source is given in 
\cite{2xtal}.  In brief, it consists of two thin, adjacent, nonlinear 
optical crystals (beta-barium-borate, BBO), cut for Type-I phase 
matching.  The crystals are aligned so that their optic axes lie in 
planes perpendicular to each other.  The pump beam and optic axis of 
the first crystal define the vertical plane, that of the second 
crystal, the horizontal plane, see Fig.~1a.  If the pump beam is 
vertically (horizontally) polarized, down conversion occurs only in 
crystal 1 (crystal 2).  When the pump polarization is set to 
45$^{\circ}$, it is equally likely to downconvert in either crystal 
\cite{notquite}.  Given the coherence and high spatial overlap between 
these two processes, the photons are created in the maximally 
entangled state $(|\rm{HH}\rangle+ e^{i \phi} |\rm{VV}\rangle) / 
\sqrt{2}$
\begin{figure}[!ht]
\begin{center}
\epsfxsize=\columnwidth
\epsfbox{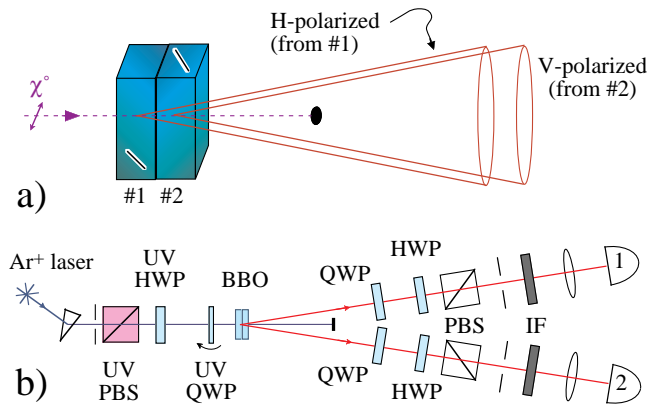}
\end{center}
\footnotesize \caption{a) Non-maximal entanglement source.  Twin 
photons are emitted along cones that originate at two identical 
down-conversion crystals, pumped by a 351 nm laser.  The crystals are 
oriented so that the optic axis of the first (second) lies in the 
vertical (horizontal) plane.  b) Experimental setup (top view) to pump 
and characterize the source.  The pump beam is wavelength and 
polarization filtered via a prism and polarizing beamsplitter (UV 
PBS), respectively.  The polarization is set by a half-wave plate (UV 
HWP).
}
\label{fig:exp}
\end{figure}
\noindent where $\phi$ is adjusted via the ultra-violet quarter-wave 
plate (UV QWP) shown in Fig.~1b.  Adjustable quarter- and half-wave 
plates (QWP \& HWP) and polarizing beamsplitters (PBS), in the two 
downconversion beams, allow polarization analysis in any basis, i.e., 
at any position on the Poincar\'{e} sphere \cite{BW}.  Each detector 
assembly comprised: an iris and a narrowband interference filter (IF, 
702 nm $\pm$ 2.5 nm), to reduce background and select (nearly-) 
degenerate photons; a 35 mm focal length lens; and a single photon 
counter (EG\&G SPCM-AQ).  The detector outputs were recorded singly, 
and in coincidence using a time to amplitude converter and a 
signal-channel analyzer.  A coincidence window of 5.27 ns was 
sufficient to capture true coincidences; since the resulting rate of 
accidental coincidences was negligible ($\sim$0.4s$^{-1}$), no 
corrections for this were necessary.

Non-maximally entangled states are produced simply by rotating the 
pump polarization.  For a polarization angle of $\chi$ with respect to 
the vertical, the output state is, $|\psi\rangle = (|\rm{HH}\rangle + 
\varepsilon e^{i \phi} |\rm{VV}\rangle)/\sqrt{1+\varepsilon^{2}}$, 
where the degree of entanglement, $\varepsilon = \rm{tan} \ \chi$ 
\cite{notquite}.  The probability of coincident detection depends on
the analyzer orientations:
\begin{eqnarray}
\rm{P}_{12}(\theta_{1},\theta_{2})
& = &
|\langle \theta_{1}| \langle \theta_{2}| \psi \rangle|^{2} \\
& = & |\rm{cos} \theta_{1} \ \rm{cos} \theta_{2} +
 \varepsilon e^{i \phi} \ \rm{sin} \theta_{1} \ \rm{sin} 
 \theta_{2}|^{2}/(1+\varepsilon^{2}) \nonumber,
\label{eq:coinc 1}
\end{eqnarray}
where for the moment we restrict ourselves to linear analyzers (i.e., 
by not using the QWP's) and $\theta_{i}$ is the orientation of the 
linear polarizer in arm $i$, with respect to the vertical.  
Traditionally, maximally entangled states are analyzed by keeping one 
analyzer fixed and varying the other; the visibility and phase of the 
resulting fringes accurately characterize the state (with this source 
we recently attained visibilities of better than 99\% \cite{2xtal}).

A related analysis method is to map out the coincidence probability 
function, in particular the distribution of the coincidence minima 
\cite{HardyExp2}.  Solving for $\rm{P}_{12}(\theta_{1},\theta_{2}) = 
0$, with $\phi=0$, we obtain $\rm{tan} \ \theta_{2} = -(\rm{cot} \ 
\theta_{1})/\epsilon$.  The shape and orientation of the resulting 
curves indicate the degree and sign of the entanglement.  
Experimentally, the coincidences are not actually zero: the value of 
the minima are a measure of the state purity.  Coincidence minima were 
found for a variety of states and analyzer settings, and fell in the 
range 0.2-1.0\%, indicating very high purities.  As Fig.~2 shows, 
across a wide range of entanglement there is good agreement between 
the experimentally determined coincidence minima and the above 
equation.  However, the data towards the bottom and the right of the 
plot are pulled off the curves slightly.  For states solely of the 
form $|$HH$\rangle$+$\varepsilon |$VV$\rangle$, this should not occur; 
however, close examination of our raw data shows there are small 
components of $|$HV$\rangle$ and $|$VH$\rangle$, even in the maximally 
entangled case.  These components become proportionally more important 
as the state becomes less entangled.  What is needed then, is an exact 
measurement of all components of the state, one that does not require 
any ana-
\begin{figure}[!ht]
\begin{center}
\epsfxsize=\columnwidth
\epsfbox{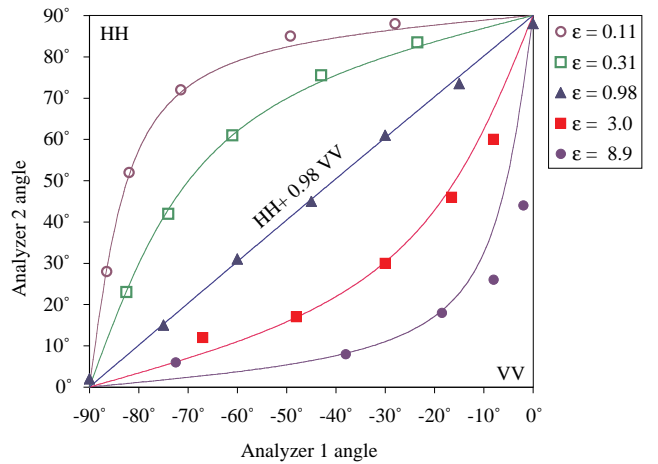}
\end{center}
\footnotesize \caption{Coincidence minima for a spectrum of non-maxim- 
ally entangled states.  {\it points\/}: experimentally determined 
coincidence minima, with uncertainties of $\pm 0.5^{\circ}$ (not 
shown); {\it curves\/}: predicted settings for zero 
coincidences, see text.}
\label{fig:minima}
\end{figure}
\noindent lytical assumptions.
 
Quantum state tomography is the solution.  For {\it continuous\/} 
variables, tomography has been implemented in both quantum 
\cite{sqztomo} and atom \cite{iontomo} optics.  Tomography is also 
possible with discrete variables \cite{discretetomo}.  For example, 
the Stokes parameters, which characterize the mean polarization of a 
classical beam \cite{Stokes}, directly yield the polarization density 
matrix for an ensemble of identically-prepared single photons 
\cite{full}.  To characterize our entangled states, we introduce the 
analogous {\it two-photon\/} Stokes parameters, which describe the 
photon polarization {\it correlations\/} in various bases.  How many 
two-photon Stokes parameters are there?  A {\it pure\/} state in 
$d$-dimensional Hilbert space is determined by 2$d$-2 linearly 
independent parameters, so only 2 and 6 parameters are respectively 
needed for the pure single and two-photon cases.  In general, however, 
the state may be partially-mixed, and $d^{2}$-1 parameters are 
required for a full analysis.  Thus 3 and 15 analyzer settings are 
respectively required to obtain the single- \cite{2vs3} and two-photon 
Stokes parameters.  In fact, we use 16 analyzer settings, as listed in 
Table~1 (the extra setting determines the normalization).  This is 
just one possible set; we will give a detailed discussion of 
discrete-variable tomography elsewhere, including extension to higher 
orders.  Briefly, we define a probability vector, {\bf P}, where the 
elements are the 16 coincidence counts normalized by the total 
coincidence rate (given by the sum of the first four measurements), 
and an invertible square matrix, {\bf M}, which is derived from the 
measurement settings.  The vector, {\bf M}$^{-1}$.{\bf P}, contains 
the real and imaginary components of the density matrix, $\hat{\rho}$.  
Elsewhere we discuss the implications of measurement drift and 
uncertainty; for the moment we simply note that there are small 
uncertainties in the final reconstructed density matrix.  The density 
matrices for a spectrum of entangled states are shown in Fig.~3 
\cite{NMRtomo}.
\begin{table}[!ht]
\caption{Settings for measuring two-photon Stokes parameters.  H, V, 
and D are respectively horizontal, vertical, and diagonal 
(45$^{\circ}$) linear polarization, L and R are left- and 
right-circular polarization.  Shown are data for the near-maximally 
entangled state of Fig.~3a (counted over 100s).}
\begin{center}
\begin{tabular}{|c|c|c||c|c|c|}
analyzer &  analyzer & coinc. & analyzer &  analyzer & coinc.  \\* 
in arm 1 & in arm 2  & count  & in arm 1 & in arm 2  & count  \\*  \hline
   H     &     H     & 34749  &    D     &     D     & 32028  \\* 
   H     &     V     &   324  &    R     &     D     & 15132  \\* 
   V     &     H     &   444  &    R     &     L     & 33586  \\* 
   V     &     V     & 35805  &    D     &     R     & 17932  \\* 
   H     &     D     & 17238  &    D     &     V     & 13441  \\* 
   H     &     L     & 16722  &    R     &     V     & 17521  \\* 
   D     &     H     & 16901  &    V     &     D     & 13171  \\* 
   R     &     H     & 16324  &    V     &     L     & 17170  \\*
\end{tabular}
\end{center}
\label{tab:settings}
%
%
\begin{figure}[!ht]
\begin{center}
\epsfxsize=\columnwidth
\epsfbox{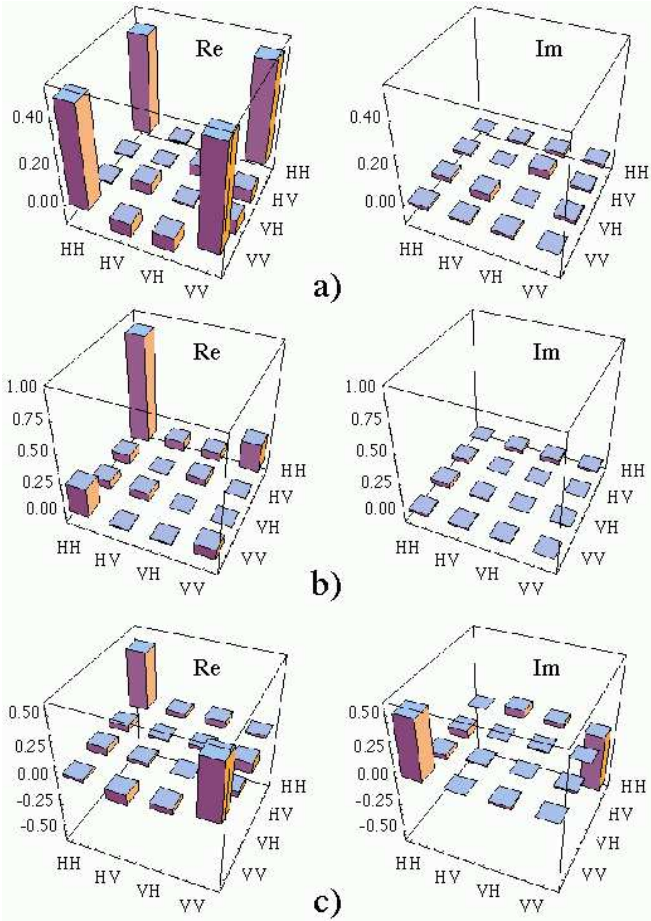}
\end{center}
\footnotesize \caption{Experimentally reconstructed density matrices 
of states that are nominally: a) $|\rm{H} \rm{H} \rangle + |\rm{V} 
\rm{V} \rangle$; b) $|\rm{H} \rm{H} \rangle + 0.3 |\rm{V} \rm{V} 
\rangle$; and c) $|\rm{H} \rm{H} \rangle - i |\rm{V} \rm{V} \rangle$.}
\label{fig:tomo}
\end{figure}
\noindent \normalsize Tomography characterizes variation both in the 
degree of entanglement, $\varepsilon$ (compare Fig.~3b with Fig.~3a), 
which is set by rotating the input polarization (with UV HWP); and in 
the phase of entanglement, $\phi$ (compare Fig.~3c with Fig.~3a), set 
by tilting the input waveplate (UV QWP).  Consider the maximally 
entangled case, Fig.~3a.   
\end{table}

\noindent Nominally, the state is 
$(|\rm{HH}\rangle+|\rm{VV}\rangle)/\sqrt{2}$ - if this were true all 
elements, except the real corner elements, would be zero.  However, as 
can be seen in the figure, some of the other elements are populated.  
What does this population signify?  From the visibilities of various 
coincidence fringes, we know the state purity is high (the visibility 
is 97.8$\pm$0.1\%), and further, mixture appears as real diagonal 
elements, so these other elements are not due to state impurity.  
(Unlike the single photon case, it is not possible to uniquely 
decompose the density matrix into pure and mixed submatrices to gain a 
direct measure of the state purity \cite{decompose}).  Instead, the 
$\sim$1\% probability of measuring $|\rm{HV}\rangle$ and 
$|\rm{VH}\rangle$ terms signifies that either the axes of the analysis 
systems were not perfectly aligned with the axes of the source 
(``horizontal'' at the analyzer is rotated with respect to 
``horizontal'' at the source), or that the optic axes of the source 
were not perfectly orthogonal, so that the produced state is, e.g., 
$|\rm{HH}\rangle$+$|\rm{V'V'}\rangle)$, where $|\rm{V'} \rangle \simeq 
|\rm{V} \rangle + \delta | \rm{H} \rangle$.

From both coincidence minima and tomography analyses, it is clear that 
we can controllably produce true non-maximally entangled states.  As 
mentioned earlier, one application of such states is testing local 
realism.  Quantum mechanics violates local realism: a quantifiable 
consequence of this, a statistical measure composed of coincidence 
measurements from a variety of analyzer settings, was first proposed 
by Bell \cite{Bell}, and has since been measured many times (see 
references in \cite{cakes,2xtal}).  All tests have found that, modulo 
some physically reasonable assumptions, nature does indeed violate 
local realism in accordance with quantum mechanics.  At some level, 
however, these measurements are unsatisfying, in that the violation is 
at a statistical level and can only be understood after some involved 
logical reasoning.  Recently Hardy proposed an ``all-or-nothing'' test 
of local realism \cite{Hardy,cakes}.  In brief, a state of the form 
$(|\rm{HH}\rangle + \varepsilon |\rm{VV}\rangle)$ is measured via four 
particular pairs of analyzer settings.  According to any local 
realistic theory, if $\rm{P}_{12}(\alpha,-\alpha) = 
\rm{P}_{12}(\beta,-\alpha^{\perp}) = 
\rm{P}_{12}(\alpha^{\perp},-\beta) = 0$ (as can be arranged by a 
suitable choice of entanglement, $\varepsilon$, and analysis angles 
$\alpha, \beta$ \cite{hardyangle}), then $\rm{P}_{12}(\beta,-\beta)=0$ 
\cite{expln}.  But quantum mechanics predicts 
$\rm{P}_{12}(\beta,-\beta)\neq0$.  More generally, 
$\rm{P}_{12}(\beta,-\beta) > \rm{P}_{12}(\alpha,-\alpha) + 
\rm{P}_{12}(\beta,-\alpha^{\perp}) + 
\rm{P}_{12}(\alpha^{\perp},-\beta)$, which is the inequality to be 
tested experimentally \cite{ineqnote}.

As Fig.~4.  shows, the ``Hardy-fraction'', 
$\rm{P}_{12}(\beta,-\beta)$, varies with degree of entanglement: for 
non- and maximally-entangled states the fraction is zero, so no test 
can be made.  The data points are normalized coincidence rates 
measured at $\pm \beta$, (to within $0.5^{\circ}$, the inherent 
uncertainty in our polarization analyzers).  The smooth curves are 
predicted directly from quantum mechanics with no adjustment 
parameters.  Clearly there is excellent agreement.  The maximum 
measured Hardy-fraction, $9.2$\% (24222 counts), occurred at an 
entanglement of $\varepsilon=0.470$; the analysis angles at this point 
were $\alpha=55.2^{\circ}$ and $\beta=72.1^{\circ}$ - the 
corresponding minima
\begin{figure}[!ht]
\begin{center}
\epsfxsize= 0.85 \columnwidth
\epsfbox{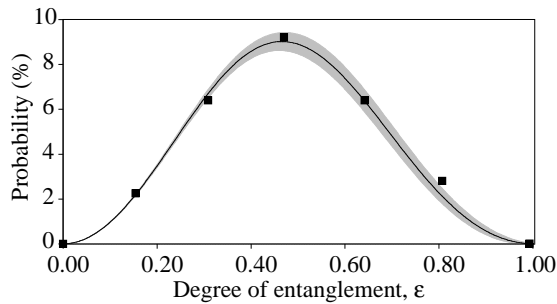}
\end{center}
\footnotesize \caption{Probability of a coincidence measurement versus 
degree of entanglement, $\varepsilon$.  {\it points\/}: measured 
probabilities obtained at $\pm (\beta \pm 0.5^{\circ})$.  
Uncertainties due to count statistics are small, the error bars lying 
within each point.  {\it curves\/}: predictions from quantum mechanics 
for analysis at exactly $\pm \beta$ (black) and out to $\pm (\beta \pm 
0.5^{\circ})$ (shaded area).}
\label{fig:hardy}
\end{figure}
\noindent were $0.43$\%, $0.52$\%, and $0.49$\% (1132, 1372, and 1279 
counts, respectively).  As the minima are determined experimentally 
the major source of error is count statistics.  Combining the above 
values, we find that the Hardy-fraction is $122 \sigma$ larger than 
the value allowed by any local realistic theory.
 
Our system should further extend access to Hilbert space, into the 
mixed regime, using depolarization technology described in 
\cite{duality}, thus allowing the production and evaluation of 
partially-mixed states of the type $|\rm{HH} \rangle \langle \rm{HH}|+ 
\gamma |\rm{VV} \rangle \langle \rm{VV}|$.  The application of mixed 
states is currently an active research area in quantum information, 
e.g., quantum secret sharing actually requires access to mixed states 
in some cases \cite{mixed}.  Finally, we note that in principle the 
photons from our source are ``hyper-entangled'', i.e., entangled in 
{\it every\/} degree of freedom, not just polarization \cite{hyper}.  
Development of a tomographic technique to measure the corresponding 
full density matrix remains a considerable challenge.

In conclusion, we have demonstrated the production, characterization, 
and utilization of true non-maximally entangled quantum states, 
without the need for postselection.  This includes a tomographic 
technique that measures the reduced density matrix for polarization 
entangled photon pairs, and a demonstration of local realism 
violation, which requires non-maximal entanglement.

We wish to thank Devang Naik for assistance, and Ulf Leonhardt, Bill 
Munro, and Mike Raymer for encouraging discussions.


\begin{references}
\bibitem{entropy} Not be confused with the {\it entropy\/} of 
entanglement [C.~H.~Bennett {\it et al.\/}, Phys.  Rev.  A {\bf 54}, 
3824 (1996)], which in this case is, $E(\varepsilon) = 
\rm{ln}(1+\varepsilon^{2}) - \varepsilon^{2} \rm{ln}(\varepsilon^{2}) 
/ (1+\varepsilon^{2}) \ $.  For $\varepsilon=1$, $E(\varepsilon)$ is a 
maximum;, for $\varepsilon=0,\infty$, $E(\varepsilon)=0$.
\bibitem{notquite} In practice, there is a slight deviation from this 
relationship as the pump beam power at the second crystal is slightly 
attenuated due to absorption in the first crystal.
\bibitem{decompose} The density matrix {\it can\/} be uniquely written 
as $\hat{\rho}=\sum_{i=1}^{\rm{N}} \lambda_{i} |\psi_{i} \rangle 
\langle \psi_{i}|$, where $\hat{\rho} |\psi_{i} \rangle = \lambda_{i} 
|\psi_{i} \rangle$ and N is the Hilbert space dimension.  However, 
this does not identify the mixed state fraction.
\bibitem{hardyangle} The Hardy angles are given by $\rm{tan} \ \alpha 
= \sqrt{1/\varepsilon}$ (where $\alpha^{\perp} = \alpha + 90^{\circ}$) 
and $\rm{tan} \ \beta = -\sqrt[3]{1/\varepsilon}$.
\end{references}
\end{document}